**Multifunctional Oxide Nanosheets: Frictional, Hall, and Piezoelectric Deformation of 2D Ga$_2$O$_3$**

*Md Akibul Islam[1], Uichang Jeong[2], Nima Barri[1], Azmeera Jannat[3], Ali Zavabeti[3], Seungbum Hong[2], Tobin Filleter[1*]*

[1] Department of Mechanical and Industrial Engineering, University of Toronto, Toronto, ON, Canada

[2] Department of Materials Science and Engineering, KAIST, Korea

[3] Department of Chemical Engineering, University of Melbourne, Parkville, Victoria, Australia

**Abstract**

Atomically thin oxides are increasingly recognized as an emerging class of 2D materials, yet their multifunctional properties have been far less investigated compared to other layered materials. Among these, gallium oxide (Ga$_2$O$_3$) is distinguished by its ultrawide bandgap, thermal stability, and mechanical rigidity, positioning it as a candidate material for nanoelectromechanical systems. In this study, the tribological, transport, and electromechanical properties of ß- Ga$_2$O$_3$ nanosheets were probed using atomic force microscopy (AFM)–based techniques. Friction force microscopy (FFM) was used to investigate interfacial sliding, and a dependence of friction on external bias was observed, which was attributed to defect-mediated charge trapping. Van der Pauw Hall measurements were conducted up to 400º C, through which the ultra wide bandgap nature of ß- Ga$_2$O$_3$ was confirmed, as electronic transport remained suppressed despite high thermal activation. Piezoresponse force microscopy (PFM) was further applied, and a measurable converse electromechanical response on the order of a few pm/V was revealed, consistent with oxygen-vacancy–induced symmetry breaking. By integrating tribological, electrical, and electromechanical measurements, it was demonstrated that ß- Ga$_2$O$_3$ nanosheets present a unique platform in which insulating stability, bias-tunable interfacial mechanics, and defect-enabled electromechanical activity coexist, offering new opportunities for multifunctional oxide nanodevices.

**Introduction**

The continued miniaturization of devices demands materials those combine mechanical durability, electrical reliability, and controlled electromechanical response within a single ultrathin platform. Two-dimensional (2D) materials are compelling in this context: atomic-scale thickness, large surface-to-volume ratios, and tunable interfaces enable functionalities that are difficult to achieve in bulk systems, spanning nanoelectronics, sensing, protective coatings, and energy transduction[1-3]. Most studied 2D crystals— graphene, MoS$_2$ and h-BN—illustrate how complementary properties (high conductivity, semiconducting behavior, and chemical robustness, respectively) can be leveraged in micro- and nanoelectromechanical systems (MEMS/NEMS)[4-6]. Yet, their performance may degrade under harsh operation: graphene oxidizes readily[7], MoS$_2$ is susceptible to humidity and high temperature[8], and h-BN, while stable, is electrically insulating in contexts requiring controlled charge transport[9]. These limitations have motivated growing interest in 2D metal oxides, an emerging class of 2D material that combines the geometry of ultrathin sheets with the robustness of ceramic oxides[10-11]. In contrast to graphene and many dichalcogenides, metal oxides are intrinsically wide-band-gap, chemically resilient, and thermally stable[12]. Such traits are attractive where minimizing leakage currents, suppressing electrostatic discharge, and preserving interfacial performance are critical, particularly in extreme environments. Their mixed ionic–covalent bonding and higher hardness also suggest improved wear resistance relative to layered chalcogenides[13]. Despite these advantages, systematic studies investigating frictional, electrical (Hall), and electromechanical (piezoelectric) responses in 2D oxides remain limited.

Among metal oxides, gallium oxide ($Ga_2O_3$) is particularly notable. In its monoclinic BETA phase-the most stable ambient structure— $Ga_2O_3$ is an ultrawide bandgap (UWBG) semiconductor ($E_g$ = 4.8 eV), with a breakdown field exceeding 8 MV/cm, far surpassing established wide-bandgap materials such as GaN and SiC.[14-15] These properties have already made $Ga_2O_3$ a leading candidate for next-generation high-voltage, high-temperature electronics[16]. Beyond electronic robustness, ß- $Ga_2O_3$ exhibits high thermal stability, resistance to oxidation, and mechanical hardness comparable to sapphire.[17] Such attributes make it uniquely suited for environments where 2D lubricants like graphene or MoS$_2$ would potentially fail. Importantly, while graphene provides conductivity and MoS$_2$ offers semiconducting functionality, $Ga_2O_3$ introduces a complementary profile: ultra wide band gap, thermally resilient and mechanically robust, enabling tribological applications where electrical insulation is also required. Recent advances in synthesis now allow $Ga_2O_3$ to be prepared in nanometer-thick forms. Techniques such as van der Waals epitaxy on layered substrates, mechanical exfoliation from bulk single crystals, and liquid-metal oxidation routes have enabled transfer or direct growth of thin $Ga_2O_3$ membranes.[18-20] Although ß-$Ga_2O_3$ do not have a naturally layered structure, these approaches demonstrate that atomically thin or few nanometer thick oxides can be isolated, opening opportunities to probe their fundamental properties.

Beyond friction and wear, $Ga_2O_3$ also presents opportunities in piezoelectric functionality. While bulk ß- $Ga_2O_3$ is centrosymmetric and nominally non-piezoelectric, certain $Ga_2O_3$ polymorphs (e.g. ε-phase) are predicted to exhibit strong piezoelectric coefficients.[21] Moreover, local symmetry breaking in ß- phase nanostructures (via surfaces, defects, or strain) may induce measurable electromechanical coupling.[22] This is particularly important in nano systems where multifunctionality is desired: a coating that not only reduces friction but also generates electrical signals under stress could simplify device architectures, enabling integrated sensing or energy harvesting. Compared with conventional piezoelectric 2D materials, $Ga_2O_3$ offers the additional benefit of high-temperature stability, ensuring reliable operation in environments that degrade typical piezoelectric dichalcogenides.[23] In

this work, we systematically investigate the tribological and piezoelectric properties of atomically thin

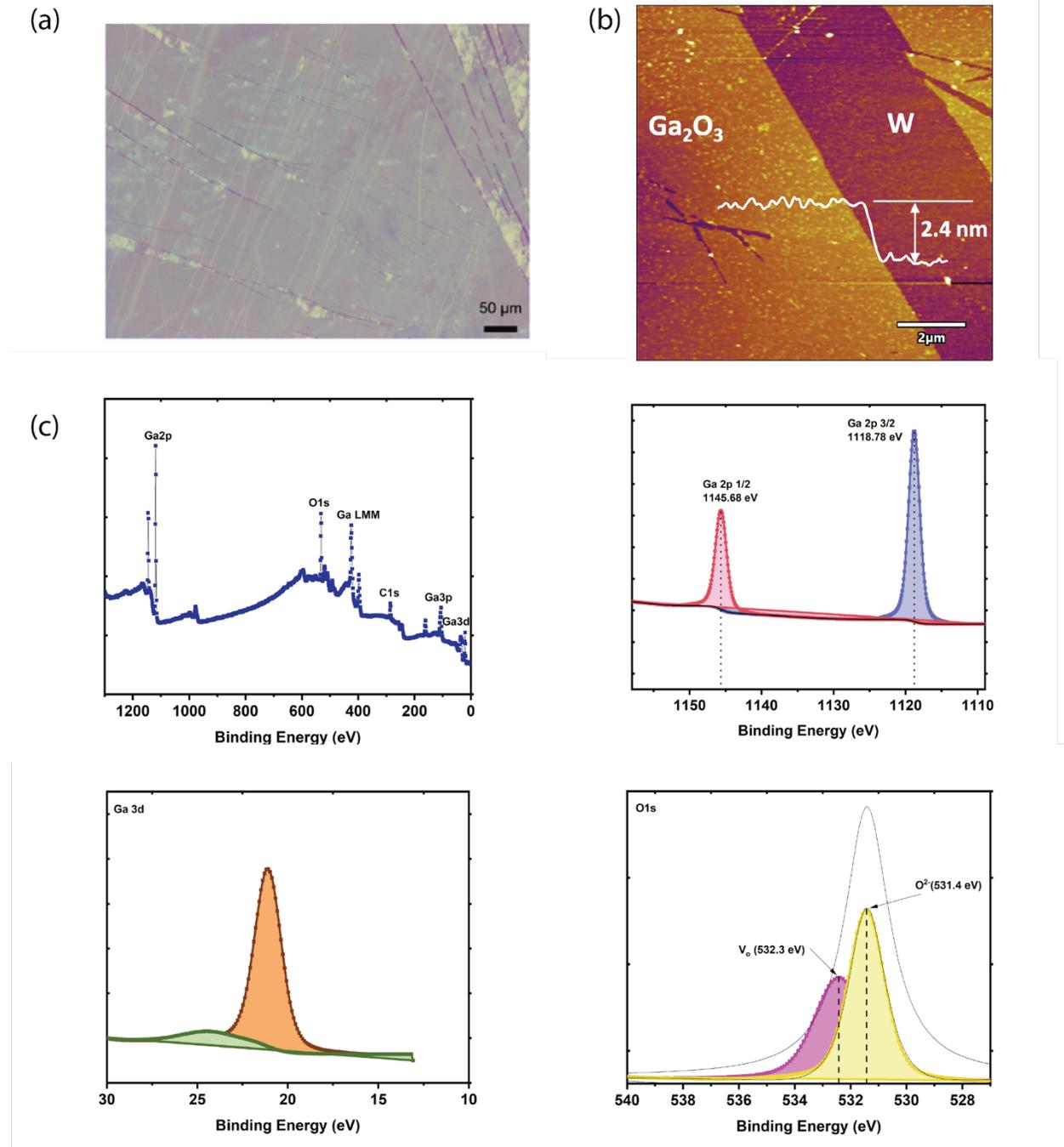

**Figure 1.** a) Optical microscopy image of exfoliated $Ga_2O_3$ flakes transferred onto a $Si/SiO_2$ substrate, showing lateral dimensions of tens of micrometers, (b) Atomic force microscopy (AFM) topography of a representative $Ga_2O_3$ flake. The height profile along the white line indicates a thickness of ~2.4 nm, confirming few-nanometer-thick layers, (c) X-ray photoelectron spectroscopy (XPS) analysis of the $Ga_2O_3$ film. The survey spectrum (top left) reveals the characteristic Ga 2p, Ga 3d, Ga 3p, and O 1s core levels, along with C 1s

contamination peak and Ga LMM Auger transition. High-resolution spectra of Ga 2p (top right) display two well-defined spin–orbit components, Ga $2p_{3/2}$ at 1118.7 eV and Ga $2p_{1/2}$ at 1145.7 eV, consistent with the +3 oxidation state of Ga. The Ga 3d peak (bottom left) centered around ~20.4 eV further confirms $Ga^{3+}$ chemical state, with a symmetric profile indicating the absence of metallic Ga. The O 1s spectrum (bottom right) is deconvoluted into two components: the main lattice oxygen peak at ~531.4 eV ($O^{2-}$ in Ga–O bonds) and a higher binding energy shoulder near 532.3 eV, attributed to oxygen vacancy–related states ($V_o$). The dominant $O^{2-}$ contribution demonstrates stoichiometric $Ga_2O_3$, while the $V_o$ peak highlights the presence of intrinsic oxygen defects.

monoclinic $Ga_2O_3$ using atomic force microscopy (AFM). Friction force microscopy (FFM) is employed to quantify nanoscale friction, adhesion, and wear behavior, while piezoresponse force microscopy (PFM) probes electromechanical response under applied bias. These complementary measurements allow direct correlation of surface mechanics with piezoelectric activity, providing a comprehensive picture of $Ga_2O_3$'s multifunctional behavior. By focusing on ß-$Ga_2O_3$, we aim to establish how its ultrawide bandgap, insulating character, and mechanically resilient lattice translate into nanoscale lubrication performance and potential piezoelectricity. Our results address a critical knowledge gap in the field of 2D metal oxides, demonstrating that ß-$Ga_2O_3$ offers a distinct combination of properties not accessible in graphene, $MoS_2$ or h-BN. Specifically, its wide bandgap ensures negligible leakage under high electric fields, its lattice hardness provides excellent wear resistance, and its potential piezoelectricity introduces active electromechanical functionality. Together, these features position 2D $Ga_2O_3$ as a promising multifunctional coating for MEMS/NEMS and flexible electronics, particularly in high-temperature and high-voltage regimes where traditional 2D lubricants are inadequate.

**Results and Discuss**

**Friction characteristic of $Ga_2O_3$**

$Ga_2O_3$ samples were prepared by directly stamping polished metal surface against a substrate to achieve large-scale exfoliation of a few layers of metal oxide as shown in Figure 1a. The optical image of the exfoliated metal oxide sheet is shown in Figure 1b. Contact mode AFM imaging was used to investigate the thickness of the exfoliated $Ga_2O_3$ and it was measured as 2.4 nm. The average roughness of $Ga_2O_3$ surface was calculated as 1.16 nm while the substrate roughness was calculated as 480.36 pm.

Friction force microscopy was used to characterize the tribological properties of $Ga_2O_3$ on tungsten (W) substrate. A cantilever tip is scanned across the sample surface under a controlled normal load, and the torsional (twisting) deflection of the cantilever is measured (Figure 2a). This torsional signal is proportional to the lateral force experienced by the tip, allowing quantitative mapping of nanoscale friction of the surface. In our study we used HQ:CSC17 tip with a normal spring constant (k) of 0.26 N/m. The tip has hard diamond like coating to prevent wear during friction experiments. The results of the friction measurements of $Ga_2O_3$ on W substrate is shown in Figure 2b. Figure 2b shows the contrast in friction between the 2D $Ga_2O_3$ sheet and the underlying substrate, where the oxide exhibits noticeably lower friction. To further explore this behavior, load-dependent friction measurements were performed, and the results are plotted in Figure 2c. $Ga_2O_3$ consistently displayed lower friction than the substrate across both low and high loads, which aligns with the general trend observed in other 2D materials. This reduction can be attributed to the atomically smooth surface of $Ga_2O_3$, which minimizes tip–surface adhesion and suppresses ploughing effects.[24] According to the lattice stick-slip theory, the rigid Ga–O bonding network produces a shallower corrugated potential landscape at the sliding interface compared to the substrate, resulting in fewer pronounced stick–slip events and reduced energy dissipation during scanning. Unlike many vdW 2D materials that exhibit tip-dependent friction due to interlayer sliding and surface puckering, atomically thin $Ga_2O_3$ (2.4 nm) did not significantly show such dependence in our measurements (Figure 2d). This behavior can be attributed to its strong ionic–covalent Ga–O bonding,[25] which yields a mechanically rigid and hard surface resistant to tip-induced deformation. One possible explanation could be, the absence of weak interlayer interactions eliminates puckering effects.

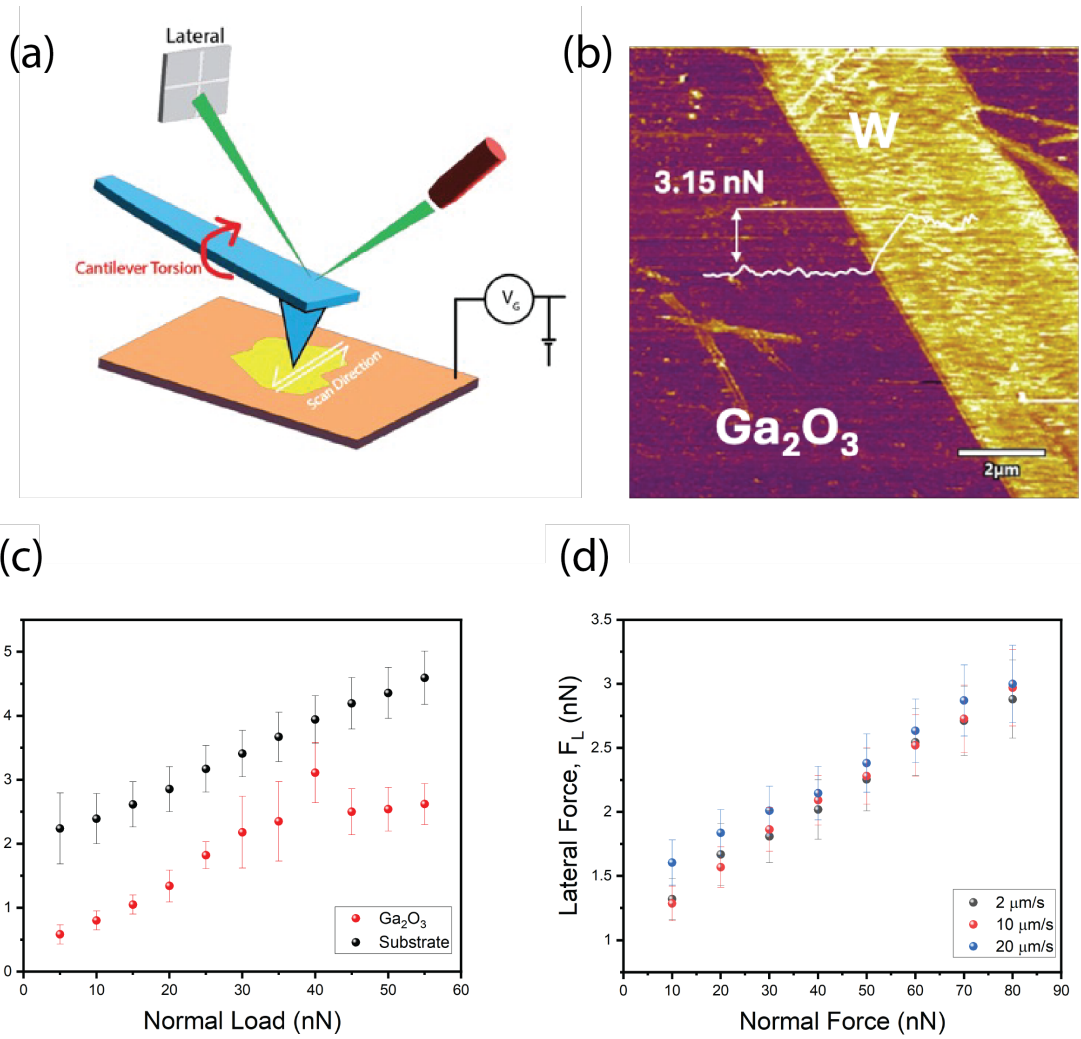

Figure 2. a) Schematic illustration of the friction force measurement set up in AFM, b) Friction force measured at 80 nN normal load. $Ga_2O_3$ (purple) showing lower friction than the substrate (yellow), c) Lateral force measured on $Ga_2O_3$ and the substrate as a function of incremental normal load, d) Lateral force measured on $Ga_2O_3$ as a function of tip sliding velocity.

In addition, the ultra-wide bandgap of $Ga_2O_3$ suppresses electronic contributions to friction, further stabilizing its response. Together, these factors render $Ga_2O_3$'s friction largely independent of tip velocity.

**Influence of trapped charge on the change of friction**

Sharp DLC (diamond like coating) coated AFM tip was used to measure friction forces and external electrical bias (DC) was applied on the substrate. In this section, we have shown that the friction of $Ga_2O_3$ can be tuned by applying an external bias voltage. To investigate the relation between the normal load and measured nanofriction at different electrical biases, a ramped normal force from 10 nN to 80 nN with intervals of 10 nN were applied under each bias voltage. Friction forces were measured between 0 to +10V and 0 to -10V. Figures 4a and 4b show the friction force change with applied bias voltage. The topography of the $Ga_2O_3$ sheet was also recorded simultaneously to investigate any change in the topography of the material during friction measurements. The scan size was 2 μm X 2 μm and the scan rate was maintained at 1 Hz and 256 line scan. Friction measurements at both positive and negative biases were conducted in the same region of the nanosheet. During both positive and negative bias applied, a linear increase of friction with the applied normal load was observed. Friction force increased from 2.77 nN at 0V to 3.47 nN at +10V (at 80 nN normal force), corresponding to a 25.3\% increase in the friction measured. When an external electric field is applied at the substrate, it leads to the polarization of the insulating tip we used in its proximity. This is the electrostatic force acting on the tip/sample interface. The increase in friction with applied bias can be further understood in terms of charge trapping at the $Ga_2O_3$ surface.[26] Under an external electric field, localized defect states and oxygen vacancies within $Ga_2O_3$ can capture and hold charges, leading to the buildup of trapped charges at the interface.[27] These trapped charges enhance the local electrostatic field and increase the adhesive interaction between the polarized AFM tip and the sample surface. As a result, the effective normal load at the sliding interface is amplified, which in turn raises the lateral resistance during scanning. This mechanism explains the observed bias-dependent increase in friction, as the accumulation of trapped charges intensifies the electrostatic contribution to the overall tip–sample interaction. Interestingly, friction increased more drastically for 0 to -10V applied bias where friction force increased from 2.77 nN at 0V to 5.41 nN at -10V (at 80 nN normal force), corresponding to a 70.5% increase in the friction. Close attention was paid to the topography of $Ga_2O_3$ to exclude the influence of surface roughness. The stronger increase of friction under negative bias can be attributed to the asymmetric trapping behavior of charges in $Ga_2O_3$. When a negative substrate bias is applied, electrons are more readily injected into the oxide due to its wide bandgap and the presence of abundant defect states such as oxygen vacancies, which act as efficient electron traps. The accumulation of these trapped negative charges at the interface substantially amplifies the local electrostatic field compared to the case of positive bias, where hole trapping is less Favorable[26]. This enhanced electrostatic attraction increases the effective tip–sample adhesion and deepens the lateral energy barriers that govern stick–slip sliding, thereby producing a much larger rise in friction. The observed 70.5\% increase in friction at −10 V, compared to the 25.3% increase at +10 V, highlights the dominant role of electron trapping in $Ga_2O_3$ during bias-modulated nano-friction. The data showing the linear increase of nano-friction is consistent with this analogy. However, under the applied electric bias fields, electrostatic adhesion between the tip and the 2D sheets must be considered. To examine the quasistatic transformation of friction, we alternated the substrate potential between 0 and −10 V while scanning the same $Ga_2O_3$ region at fixed load (Fig. 4c). The resulting two-level friction signal is consistent with bias-controlled trapping and detrapping of charge in near-surface defect/vacancy states of $Ga_2O_3$. At −10 V, electrons are injected and captured by these states, which increases the local surface potential, strengthens polarization of the insulating tip, and thereby raises the electrostatic (adhesive) contribution to the effective normal load; the higher effective load manifests as a higher friction plateau. Returning to 0 V reduces charge injection and allows trapped electrons to relax/neutralize, diminishing the interfacial field and restoring the lower friction plateau. The fact that the friction toggles reproducibly over three cycles and remains stable over ~40 s intervals, while the topography is unchanged, indicates that the modulation arises from reversible electrostatic interactions governed by trap occupancy and dielectric relaxation (i.e., the RC timescale of the tip–oxide–substrate stack), rather than from wear or morphological changes. This cyclic behavior coherently explains the stronger friction under negative bias and its return toward the 0 V baseline when the field is removed.

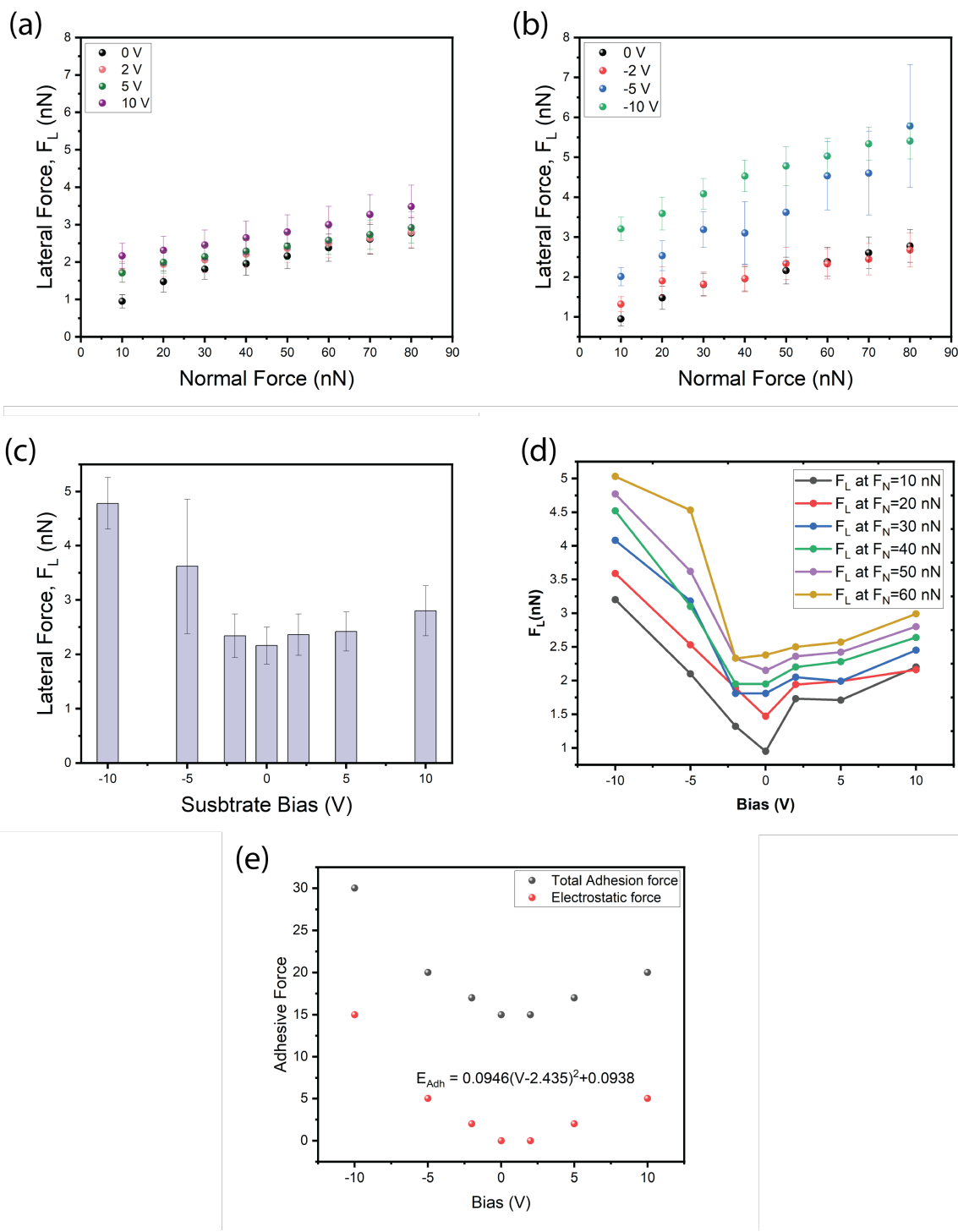

**Figure 3.** Lateral force measured by friction force micriscopy as a function of normal force for (a)positive and (b)negative bias applied at the substrate, (c) Lateral force plotted as a function of substrate bias at 50 nN normal load, (d)Lateral force plotted as a function of substrate bias at all normal loads, (e) Calculation for electrostatic force as a function of substrate bias.

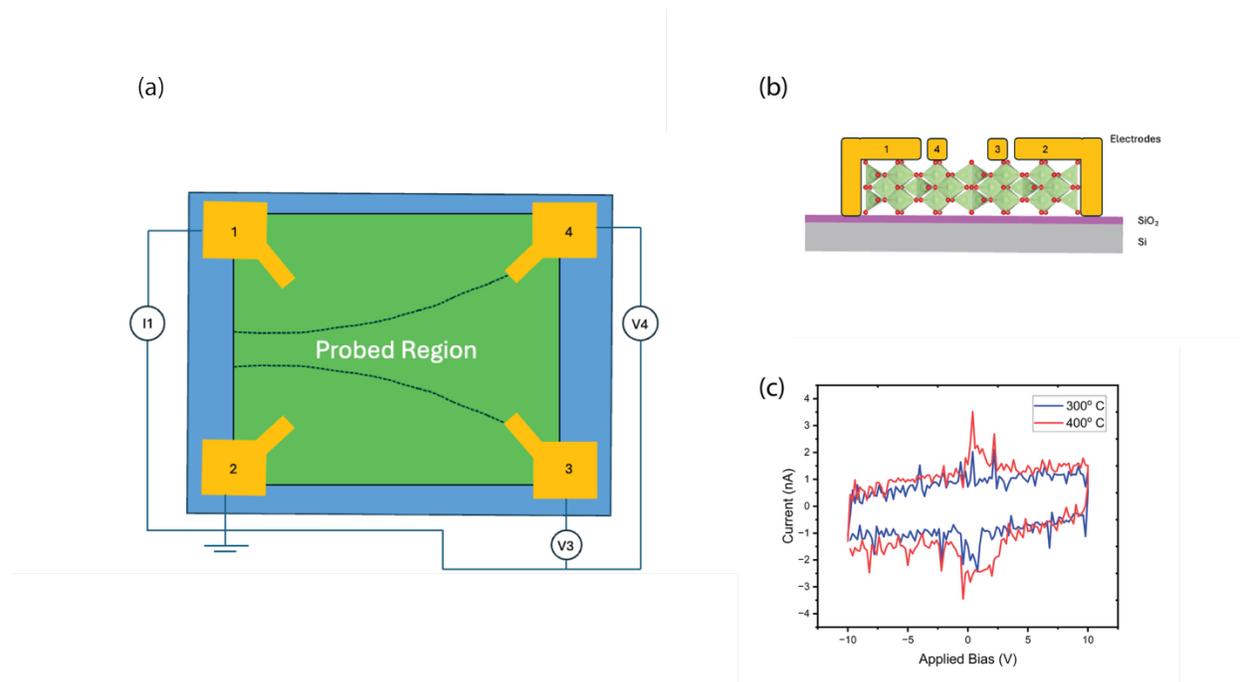

**Figure 4.** a) Schematic top-view layout of the van der Pauw Hall device fabricated from exfoliated $Ga_2O_3$ nanosheets on a 280 nm $SiO_2$/Si substrate. Four Cr/Au electrodes (1–4) define the probing geometry, enabling current injection ($I_1$) and voltage detection ($V_3$, $V_4$) across the central $Ga_2O_3$ region. (b) Cross-sectional schematic of the device, showing the $Ga_2O_3$ nanosheet contacted by electrodes and supported on the $SiO_2$/Si substrate. (c) Current–voltage (I–V) characteristics of the device measured under a voltage sweep from –10 V to +10 V at elevated temperatures of 300 °C (blue) and 400 °C (red). The negligible current response indicates insulating behavior of $Ga_2O_3$ even under high-temperature biasing, consistent with its ultrawide bandgap and low intrinsic carrier density.

**Tuning friction of $Ga_2O_3$ under electrical bias**

Figure 4d demonstrates the change of nanofriction of $Ga_2O_3$ from -10V to +10V at 50 nN normal load. Careful observation can reveal that the change of apex of the bars follow a parabolic curve which is consistent with previous friction measurements of 2D materials using insulating tip (Dynamically tuning friction at the graphene interface using the field effect). Friction force measured scales parabolically with the applied bias as $F_L \propto V$. By comparing the adhesion between the AFM tip and the 2D sheet before and after the application of bias voltage, electrostatic adhesion between the AFM tip and 2D sheet can easily be calculated. We plotted the contribution of electrostatic adhesion at each bias voltage (Figure 4f) and by fitting the curve, we formulated an equation for tuning the electrostatic adhesion between the tip and sample at any bias voltage:

$$E_{Adh} = 0.0964(V - 2.435)^2 + 0.0938$$

**Negligible Hall Mobility of $Ga_2O_3$ at Elevated Temperature**

The $Ga_2O_3$ device was fabricated in a van der Pauw configuration with Cr/Au contacts and measured on a temperature-controlled stage at 100 to 400 °C at 50 °C intervals. At each setpoint, Hall voltage and current–voltage sweeps were recorded after sufficient thermal stabilization. Across all conditions, the device exhibited open-circuit behavior, with negligible current response and no measurable Hall mobility. This result reflects the intrinsic limitations of ß-$Ga_2O_3$ as a charge transport medium in its undoped, thin-film form. The ultrawide bandgap of ~4.8–4.9 eV restricts thermal excitation of carriers, so the intrinsic electron concentration remains extremely low even at 400 °C, far below the threshold required for Hall detection. Consequently, the conduction band remains largely unpopulated, resulting in vanishing carrier density. Electron mobility in ß-$Ga_2O_3$, while adequate in doped bulk crystals or epitaxial films (10–300 cm$^2$/V·s depending on growth orientation and defect density), is inherently limited by the monoclinic lattice symmetry. At elevated temperatures, strong polar optical phonon scattering dominates, further reducing mobility and making charge transport increasingly inefficient. Ionized impurity scattering and defect-related localization add to this suppression, ensuring that any thermally activated carriers have very short mean free paths. The Cr/Au- $Ga_2O_3$ contact interfaces also play a critical role: large Schottky barriers impede carrier injection, preventing the electrodes from supplying sufficient free electrons for measurable conduction. Moreover, intrinsic point defects such as oxygen vacancies, which are abundant in $Ga_2O_3$, act as deep traps. These states capture injected or thermally generated carriers, screen the applied field, and pin the Fermi level near midgap, thereby eliminating the possibility of gate- or bias-induced carrier modulation.

The combination of ultrawide bandgap–driven carrier scarcity, temperature-dependent phonon-limited mobility, defect-mediated charge trapping, and Schottky-barrier-limited injection explains the absence of Hall mobility and the observed open-circuit behavior across the studied temperature range. These findings underscore that while ß-$Ga_2O_3$ is unsuitable for conventional charge transport in undoped thin-film form, its suppressed conduction and high thermal stability enhance its potential as a multifunctional oxide. In particular, the negligible free-carrier density minimizes electronic screening, allowing subtle electromechanical effects such as piezoelectricity to be probed with higher sensitivity. Thus, the insulating character that hinders electronic conduction simultaneously provides an opportunity to explore and harness the piezoelectric response of ß-$Ga_2O_3$ at the nanoscale.

**Piezoelectricity of 2D $Ga_2O_3$**

The piezoresponse of $Ga_2O_3$ nanosheets was investigated using piezoresponse force microscopy (PFM), in which an AC voltage was applied to the sample surface through the conductive tip to induce electromechanical displacement (Figure 5a). Specifically, we utilized dual AC resonance tracking PFM (DART-PFM), which continuously tracks and adjusts the excitation frequency to the local contact resonance of the tip–sample system during scanning. The topography of the scanned region is shown in Figure 5b, with the corresponding PFM phase and amplitude images (Figures 5c and 5d) confirming a spatially uniform piezoresponse across the nanosheets. To determine the effective piezoelectric coefficient, $d_{33}^{eff}$, the drive AC voltage was first set to 1 V and gradually increased to 4 V with increment of 1 V in PFM measurements. The measured amplitude, representing the tip-induced surface displacement, exhibited a linear dependence on the applied bias, yielding a $d_{33}^{eff}$ value of 5.39 pm/V from the slope of the amplitude–voltage plot (Figure 5e).

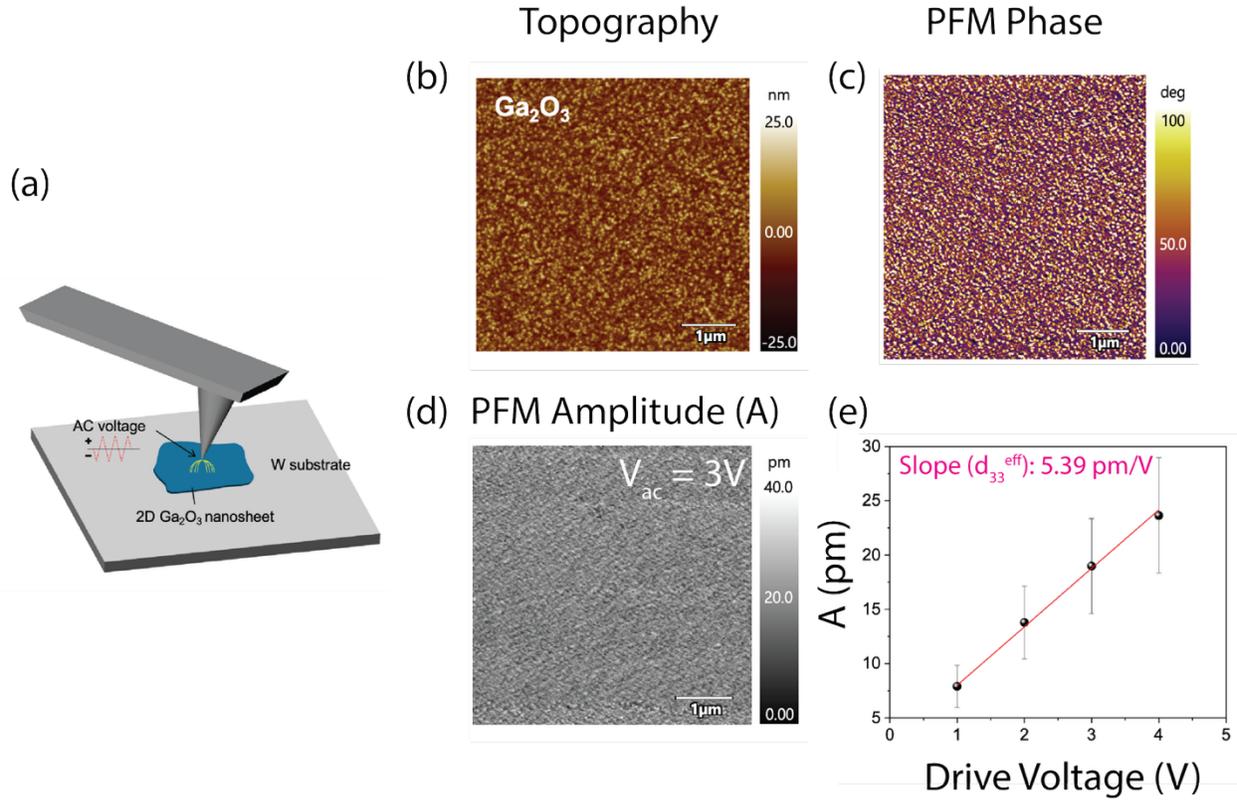

**Figure 5.** DART-PFM measurements and determination of effective piezoelectric coefficient, $d_{33}^{eff}$. (a) Schematic illustration of PFM measurements. (b) Topography and (c) corresponding PFM phase and (d) PFM amplitude images of 2D Ga$_2$O$_3$ nanosheets (V$_{ac}$=3 V). (e) Linear plot of drive voltage (V$_{ac}$) vs. PFM amplitude, of which slope corresponds to $d_{33}^{eff}$

The PFM phase showed clear contrast between W substrate and Ga$_2$O$_3$ region. Notably, the substrate region displayed higher amplitudes than Ga$_2$O$_3$ region with all the applied drive voltage, which, although showing a linear dependence, can be strongly attributed to the capacitive force-induced artifacts on the non-piezoelectric materials. The amplitude on the Ga$_2$O$_3$ region in the same images exhibited linear correlation resulting in similar $d_{33}^{eff}$ value of 4.27 pm/V in agreement with the result from Figure 5e. Among the polymorphs of Ga$_2$O$_3$, only the ε-phase is intrinsically piezoelectric, with reported coefficients of ~10–11 pm/V4. Therefore, the piezoresponse observed in our ß-Ga$_2$O$_3$ nanosheets is likely to be associated with oxygen vacancy-induced breaking of centrosymmetry, which enables electromechanical coupling under an external bias.

To further evaluate whether the ß-Ga$_2$O$_3$ nanosheets exhibit ferroelectric-like switchable polarization, we carried out DART switching spectroscopy PFM (DART-SS-PFM) measurements. In SS-PFM, a DC bias waveform is applied locally while the tip monitors the electromechanical response, producing amplitude and phase hysteresis loops. In our measurements, the ß-Ga$_2$O$_3$ nanosheets occasionally displayed a weak hysteresis, the phase exhibited a small lagging switch with a narrow coercive voltage, and the corresponding amplitude showed a slight "butterfly" feature near the coercive biases. Such behavior suggests a non-linear and partially reversible component in the nanosheet response. However, we found that these hysteresis loops were not consistently reproducible across repeated measurements or different regions. The lack of reproducibility, combined with the narrow coercive voltage window and the absence of a sizable remanent piezoresponse, strongly indicates that the loops originate from extrinsic effects rather than true ferroelectric switching. Similar pseudo-ferroelectric signals are widely reported in the PFM literature, where they are often attributed to mechanisms such as charge injection, electrostatic interactions, or field-induced migration of oxygen vacancies.[28] Given that $Ga_2O_3$ nanosheets showed oxygen-vacancy–related states in XPS and were insulating, it is plausible that such defects or injected charges under bias contribute to the observed hysteresis-like response. Therefore, while the nanosheets clearly exhibited a defect-mediated electromechanical coupling, the SS-PFM data did not provide sufficient evidence for robust or switchable ferroelectricity. Additional experiments, such as non-volatile domain writing or macroscopic P–E measurements, would be required to make such a claim.

**Conclusion**

In this study, we investigated the multifunctional behavior of atomically thin ß-$Ga_2O_3$ through frictional, electronic, and electromechanical characterization. Friction force microscopy showed that $Ga_2O_3$ exhibits lower friction than the underlying substrate and that friction can be tuned under an applied bias, with stronger modulation observed under negative bias. This bias dependence is consistent with charge trapping at oxygen-vacancy–related sites, indicating that surface defects provide an opportunity to control nanoscale tribological behavior. Hall effect measurements in a van der Pauw geometry revealed negligible current response across 100–400°C, demonstrating that ß-$Ga_2O_3$ remains insulating and exhibits no measurable Hall mobility even at elevated temperatures. The absence of conduction highlights the intrinsic limitation imposed by its ultrawide bandgap and strong phonon scattering, which restrict electronic transport and suppress free-carrier contributions to interfacial processes. Thus, while high-temperature electronic activity remains minimal, external bias still provides an effective route to modify interfacial friction, underscoring a decoupling between bulk transport and surface tribological response. Piezoresponse force microscopy further confirmed that ß-$Ga_2O_3$ nanosheets exhibit a reproducible electromechanical displacement with an effective $d_{33}^{eff}$ of ~4–5 pm/V. This response is attributed to oxygen-vacancy–induced symmetry breaking, as supported by XPS evidence of defect-related $O^{1s}$ states. Occasional hysteresis loops observed in switching spectroscopy PFM were not consistently reproducible and are attributed to extrinsic effects such as charge injection rather than robust ferroelectric switching. Together, these results establish that atomically thin ß-$Ga_2O_3$ simultaneously offers insulating stability at elevated temperatures, bias-tunable frictional properties, and defect-mediated electromechanical activity, making it a promising multifunctional oxide for applications in MEMS/NEMS and flexible electronics.